\documentclass[prb,aps,reprint,superscriptaddress]{revtex4-1}
\usepackage{graphics}
\usepackage{inputenc}
\usepackage{epsfig}
\usepackage{amsmath}
\usepackage{amsfonts}
\usepackage{amssymb}
\usepackage{graphicx}
\usepackage{color}
\usepackage{tabularx}
\usepackage{txfonts}

\newcommand\st{\bgroup\markoverwith
  {\textcolor{blue}{\rule[.35ex]{5pt}{1.1pt}}}\ULon}

\begin{document}

\title{Pronounced 2/3 magnetization plateau in a frustrated $S$ = 1 isolated spin-triangle compound: Interplay between Heisenberg and biquadratic exchange interactions}

\author{S. Chattopadhyay}
\email{s.chattopadhyay@hzdr.de}
\affiliation{Dresden High Magnetic Field Laboratory (HLD-EMFL), Helmholtz-Zentrum Dresden-Rossendorf, 01328 Dresden, Germany}

\author{B. Lenz}
\email{benjamin.lenz@polytechnique.edu}
\affiliation{CPHT, CNRS, Ecole Polytechnique, IP Paris, F-91128 Palaiseau, France}
\author{S. Kanungo}
\affiliation{School of Physical Sciences, Indian Institute of Technology Goa, 403401 Goa, India}

\author {Sushila}
\affiliation{Department of Chemistry and Centre for Advanced Studies in Chemistry, Panjab University, Chandigarh 160014, India}

\author{S. K. Panda}
\affiliation{CPHT, CNRS, Ecole Polytechnique, IP Paris, F-91128 Palaiseau, France}
\affiliation{Department of Physics, Bennett University, Greater Noida 201310, Uttar Pradesh, India}

\author{S. Biermann}
\affiliation{CPHT, CNRS, Ecole Polytechnique, IP Paris, F-91128 Palaiseau, France}
\affiliation{Coll{\`e}ge de France, 11 place Marcelin Berthelot, 75005 Paris, France}

\author{W. Schnelle}
\affiliation{Max Planck Institute for Chemical Physics of Solids, N\"othnitzer Stra\ss e 40, 01187 Dresden, Germany}

\author{K. Manna}
\affiliation{Max Planck Institute for Chemical Physics of Solids, N\"othnitzer Stra\ss e 40, 01187 Dresden, Germany}

\author{R. Kataria}
\affiliation{Department of Chemistry and Centre for Advanced Studies in Chemistry, Panjab University, Chandigarh 160014, India}

\author{M. Uhlarz}
\affiliation{Dresden High Magnetic Field Laboratory (HLD-EMFL), Helmholtz-Zentrum Dresden-Rossendorf, 01328 Dresden, Germany}

\author{Y. Skourski}
\affiliation{Dresden High Magnetic Field Laboratory (HLD-EMFL), Helmholtz-Zentrum Dresden-Rossendorf, 01328 Dresden, Germany}

\author{S. A. Zvyagin}
\affiliation{Dresden High Magnetic Field Laboratory (HLD-EMFL), Helmholtz-Zentrum Dresden-Rossendorf, 01328 Dresden, Germany}

\author{A. Ponomaryov}
\affiliation{Dresden High Magnetic Field Laboratory (HLD-EMFL), Helmholtz-Zentrum Dresden-Rossendorf, 01328 Dresden, Germany}

\author{T. Herrmannsd\"orfer}
\affiliation{Dresden High Magnetic Field Laboratory (HLD-EMFL), Helmholtz-Zentrum Dresden-Rossendorf, 01328 Dresden, Germany}

\author{R. Patra}
\affiliation{Department of Chemistry and Centre for Advanced Studies in Chemistry, Panjab University, Chandigarh 160014, India} 

\author{J. Wosnitza}
\affiliation{Dresden High Magnetic Field Laboratory (HLD-EMFL), Helmholtz-Zentrum Dresden-Rossendorf, 01328 Dresden, Germany}
\affiliation{Institut f\"ur Festk\"orper- und Materialphysik, Technische Universit\"at Dresden, 01062 Dresden, Germany}

\date{\today}

\begin{abstract}
{We report the synthesis and characterization of a new quantum magnet [2-[Bis(2-hydroxybenzyl)aminomethyl]pyridine]Ni(II)-trimer (BHAP-Ni$_3$) in single-crystalline form. Our combined experimental and theoretical investigations reveal an exotic spin state that stabilizes a robust 2/3 magnetization plateau between 7 and 20 T in an external magnetic field. AC-susceptibility measurements show the absence of any magnetic order/glassy state down to 60 mK. The magnetic ground state is disordered and specific-heat measurements reveal the gapped nature of the spin excitations. Most interestingly, our theoretical modeling suggests that the 2/3 magnetization plateau emerges due to the interplay between antiferromagnetic Heisenberg and biquadratic exchange interactions within nearly isolated spin $S$ = 1 triangles.}
\end{abstract}

\maketitle{}
\section{Introduction}
A spin triangle with nearest-neighbor antiferromagnetic (AFM) interactions is the most fundamental building block to comprehend consequences of magnetic frustration~\cite{moe06,lac11,diep13}. As the archetypical model for investigating frustration-driven quantum behaviors, efforts have been made to explore the novel and exotic magnetic phase diagram in frustrated triangular spin systems both from theoretical and experimental perspectives. Those efforts have unfolded the existence of unusual and intricate spin states of quantum origin. One of such interesting phenomena is the existence of fractional magnetization plateaus in the field dependence of magnetization. Such plateaus are instigated by quantum fluctuations which promote the lifting of a continuous ground-state degeneracy in an antiferromagnetic Heisenberg triangular spin system~\cite{lac11,chu91,zhi00,hen89,kam18}. 

\par
In many cases, the appearance of a 1/3 magnetization plateau has been observed. Chubukov et al. showed that in a two-dimensional (2D) AFM triangular system, quantum fluctuations favor an up-up-down type collinear spin arrangement leading to a 1/3 plateau  in presence of a magnetic field ($H$) where the net magnetization is 1/3 of the fully polarized state ~\cite{chu91}. So far, 1/3 plateaus have been observed in frustrated triangular systems such as C$_6$Eu, Cs$_2$CuBr$_4$, RbFe(MoO$_4$)$_2$, Rb$_4$Mn(MoO$_4$)$_3$, CsFe(SO$_4$)$_2$, GdPd$_2$Al$_3$, Ba$_3$NiSb$_2$O$_9$, and Ba$_3$CoSb$_2$O$_9$~\cite{sue81,ono03,ina96,ish11,kit99,shi12,shi11,svi03}. There is also one example (a \{Cu$_3$-As\}-type triangular spin ring), where a  1/2 plateau was observed in a triangular antiferromagnet ~\cite{cho06}. Other metal-organic frameworks including triangular Ni rings were found to have ferromagnetic coupling between Ni atoms and, hence, show no fractional magnetization plateau at all \cite{Schmitz18,Zhang09}. 

\par
In this work, we report on the magnetic properties of a new frustrated metal-organic system Ni$_3$O$_6$N$_6$C$_{60}$H$_{54}$ ([2-[Bis(2-hydroxybenzyl)aminomethyl]pyridine]Ni(II)-trimer; abbreviated as BHAP-Ni$_3$ below). Contrary to other 2D/layered systems with antiferromagnetic spin-spin couplings, BHAP-Ni$_3$ provides an excellent opportunity to investigate the physics of a fundamental frustrated spin-triangle-unit as it consists of Ni$^{2+}$ ($S$ = 1) triangles with each triangular unit being essentially magnetically decoupled from the others due to the large intertriangle Ni-Ni distances (Fig. 1). 
In addition, the small spin value of Ni ($S$ = 1) is suitable in preserving the quantum character of the magnetism involved. 

\begin{figure}[bth]
	\includegraphics [width=1.0\linewidth, angle=0]{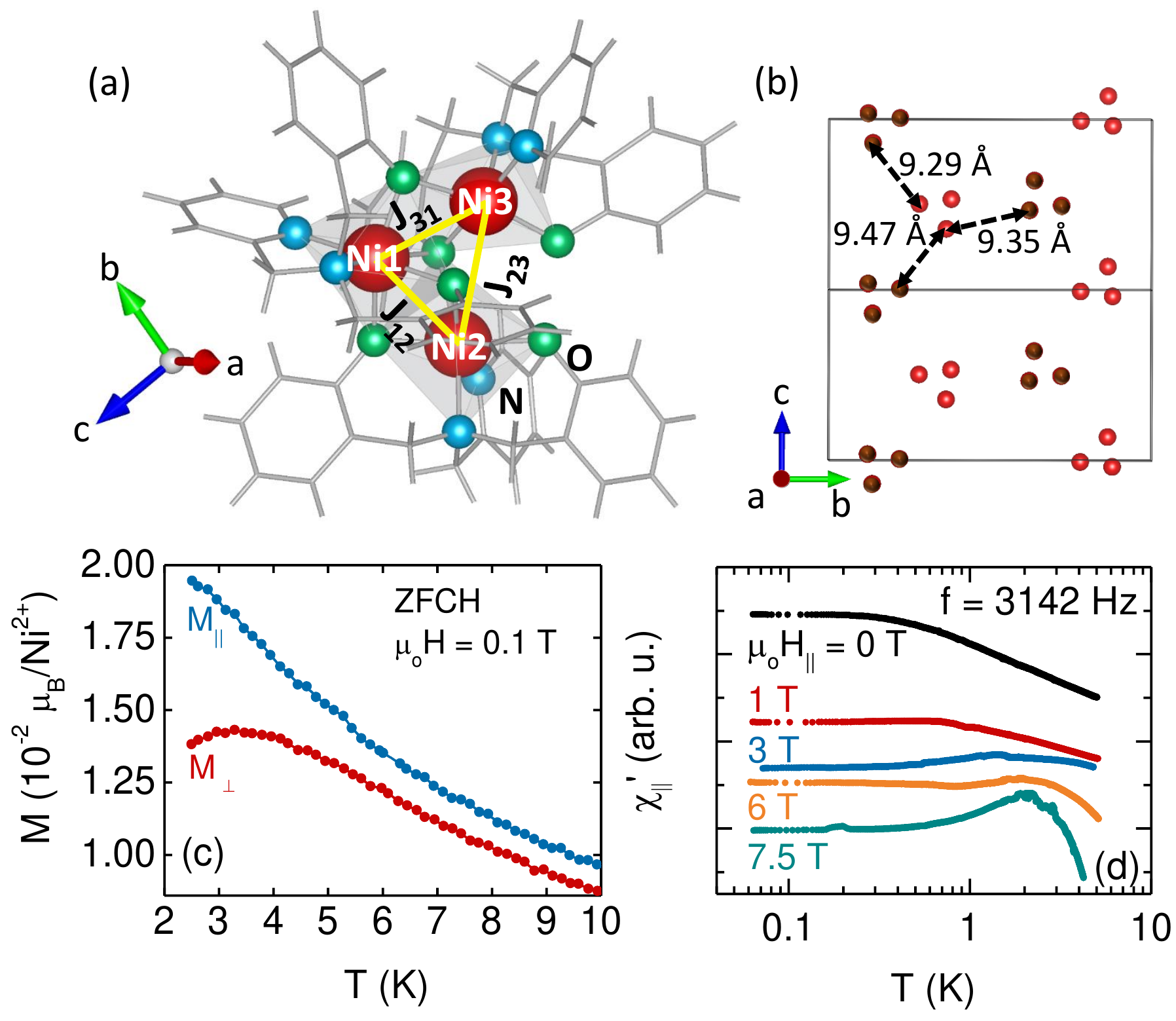}
	\caption{(a) A perspective view of the Ni$^{2+}$ (red spheres) spin triangle along with the exchange interactions (yellow lines) in BHAP-Ni$_3$. Each Ni ion is situated in a distorted octahedral environment (gray shadow) formed by four O (green spheres) and two N (light blue spheres) ions. (b) Distribution of isolated spin triangles in the lattice, light (dark) red circles denote Ni ions in the lower (upper) layer along the $a$ direction. (c) Zero-field-cooled magnetization as a function of temperature. The magnetic field was applied both parallel ($M_\parallel$) and perpendicular ($M_\perp$) to the triangle plane. (d) Temperature dependence of the real part of the ac susceptibility measured without and at several magnetic fields applied within the triangle plane ($H_\parallel$).}
	\label{Fig1}
\end{figure}   

\par
Our investigation reveals that BHAP-Ni$_3$ has a magnetically disordered $S^{\mathrm{tot}}\hspace{-.15cm}=\hspace{-.15cm}0$ ground state. 
Under the influence of a magnetic field, this system reaches a quasi-stable state $\vert S^z\hspace{-.15cm}=\hspace{-.15cm}1,S^z\hspace{-.15cm}=\hspace{-.15cm}1,S^z\hspace{-.15cm}=\hspace{-.15cm}-1\rangle$ with two $S\hspace{-.15cm}=\hspace{-.15cm}1$ quantum spins in the $S^z\hspace{-.15cm}=\hspace{-.15cm}1$ and one in the $S^z\hspace{-.15cm}=\hspace{-.15cm}-1$ configuration\footnote{Note that this notation just specifies the manifold of microstates that form the ground state. Still all permutations of a given spin combination contribute to it as discussed in context of Fig.~\ref{FigN1}(a).} producing a very feeble anomaly near 1/3 of the saturation magnetization ($M_\mathrm{S}$). 
With further increase in field,  a non-trivial $S^{\mathrm{tot}}\hspace{-.15cm}=\hspace{-.15cm}2$ state ($\vert1,1,0\rangle$) emerges, where two spins are aligned along the field-direction and one spin is in its $S^z\hspace{-.15cm}=\hspace{-.15cm}0$ configuration. This state stabilizes a 2/3 plateau in the magnetization curve, which spans a wide range of field. To our knowledge, such a clear and robust 2/3 plateau has not been realized so far. The only example could be Cs$_2$CuBr$_4$, where only a weak anomaly was visible near $\frac{2}{3}M_\mathrm{S}$~\cite{ono03,ono11}. Theoretical modeling of the system by anisotropic antiferromagnetic exchange interactions and single-ion anisotropy cannot account for this 2/3 plateau. It turns out that including biquadratic exchange between the Ni atoms in the modeling of the system is essential to explain its magnetic behavior at large magnetic fields. Therefore, BHAP-Ni$_3$ is one of the rare physical systems where biquadratic exchange drives the magnetic properties and could serve as a model to study the interplay of quadrupolar and dipolar exchange.

\section{Synthesis and Characterization} 
The single crystals of BHAP-Ni$_3$ were grown using a wet-chemical synthesis technique. A room-temperature single-crystal x-ray diffraction study shows that  BHAP-Ni$_3$ crystallizes in a monoclinic structure (space group $P2_1/n$) with lattice parameters $a$ = 12.9067 $\AA$, $b$ = 29.1872 $\AA$, $c$ = 15.4992 $\AA$, and $\beta$ = 96.86$^\circ$. The unit cell is composed of Ni-triangle units where the Ni-Ni bond lengths are 2.87, 2.92, and 3.42 $\AA$, respectively. These triangles are quite isolated as the shortest distance between the two Ni ions of two different triangular units is 9.29 $\AA$, see Fig. 1(b). More details of the synthesis and subsequent structural analysis are provided in Appendices~\ref{App:Sample} and \ref{App:XRD}.

\section{Results and Discussion}
DC-magnetization ($M$) measurements up to 14 T were performed under zero-field cooled protocol in a vibrating-sample magnetometer. The ac-susceptibility measurements were carried out in a dilution refrigerator down to 60 mK. The high-field magnetization was studied up to 40 T using a pulsed-field magnet equipped with a $^3$He cryostat. Specific-heat measurements were performed down to 360 mK by means of the relaxation method (HC option, PPMS9, Quantum Design). Using electron spin resonance (ESR) measurements, the paramagnetic $g$-factor was extracted from the linear fit of the frequency-field dependence, revealing $g$ = 2.23. 

\par 
To calculate the magnetic properties of BHAP-Ni$_3$, we constructed a spin model for an isolated Ni triangle, which involves strongly asymmetric AFM exchange terms between the nickel spins. The model was solved using exact diagonalization to calculate the magnetization, magnetic susceptibility, and specific heat. A direct comparison between theory and experiment unravels the significance of various terms in our spin-Hamiltonian and allows a microscopic interpretation of our experimental results. Most interestingly, we find that including a biquadratic exchange term is essential to reproduce both the experimental magnetization curve and the measured specific heat curves.

\par
DC magnetization of BHAP-Ni$_3$ was measured as a function of temperature ($T$) with the magnetic field ($H$) applied both in the plane of the triangles ($M_\parallel$) and perpendicular ($M_\perp$) to it. As shown in Fig.~\ref{Fig1}(c), there is no signature of any sharp anomaly down to 2 K indicating the absence of long-range magnetic order. $M_\parallel$ ($T$) is slightly larger than $M_\perp$ showing presence of a weak easy-plane type anisotropy. However, no in-plane anisotropy was observed in the measurement (Fig.~\ref{FigS2} in the appendix). Curie-Weiss-type fitting to the susceptibility data measured in a powder-sample in the paramagnetic region (above 100 K) resulted a Weiss temperature of about -8 K in BHAP-Ni$_3$ indicating a dominance of AFM coupling (see Fig.~\ref{FigS2} in the appendix). 

\par
Figure \ref{Fig1}(d) shows the temperature dependence of the real part of the ac magnetic susceptibility ($\chi^\prime$) measured down to 60 mK with an excitation frequency ($f$) of 3.142 kHz. In zero dc-field, absence of any sharp anomaly in $\chi^\prime$ down to the lowest measured temperature is a signature of the absence of long-range magnetic order. By cooling below 4.2 K in 0 T, $\chi^\prime$ increases and approaches a temperature-independent flat region below $\sim$400 mK. The extent of this flat region was found to be $f$ independent ruling out any possibility of spin freezing as well. 
A flat region in $\chi^\prime$ implies a linear temperature dependence of the magnetization. At temperatures low compared to the AFM exchange energy involved, such a behavior is presumably a signature of an intra-triangle spin disordered state where quantum fluctuations prevail due to frustration. $\chi^\prime$ was also measured in presence of dc magnetic fields. With increasing dc field, the shape of the $\chi^\prime$($T$) curve changes and at $\mu_0H$ = 3 T, a broad maximum starts to emerge  around 2 K as can be seen from Fig. \ref{Fig1}(d). This feature gets more prominent with increasing dc field indicating the onset of a field-induced transition. 

\par
\begin{figure}[htbp]
	\includegraphics [width=0.99\linewidth, angle=0]{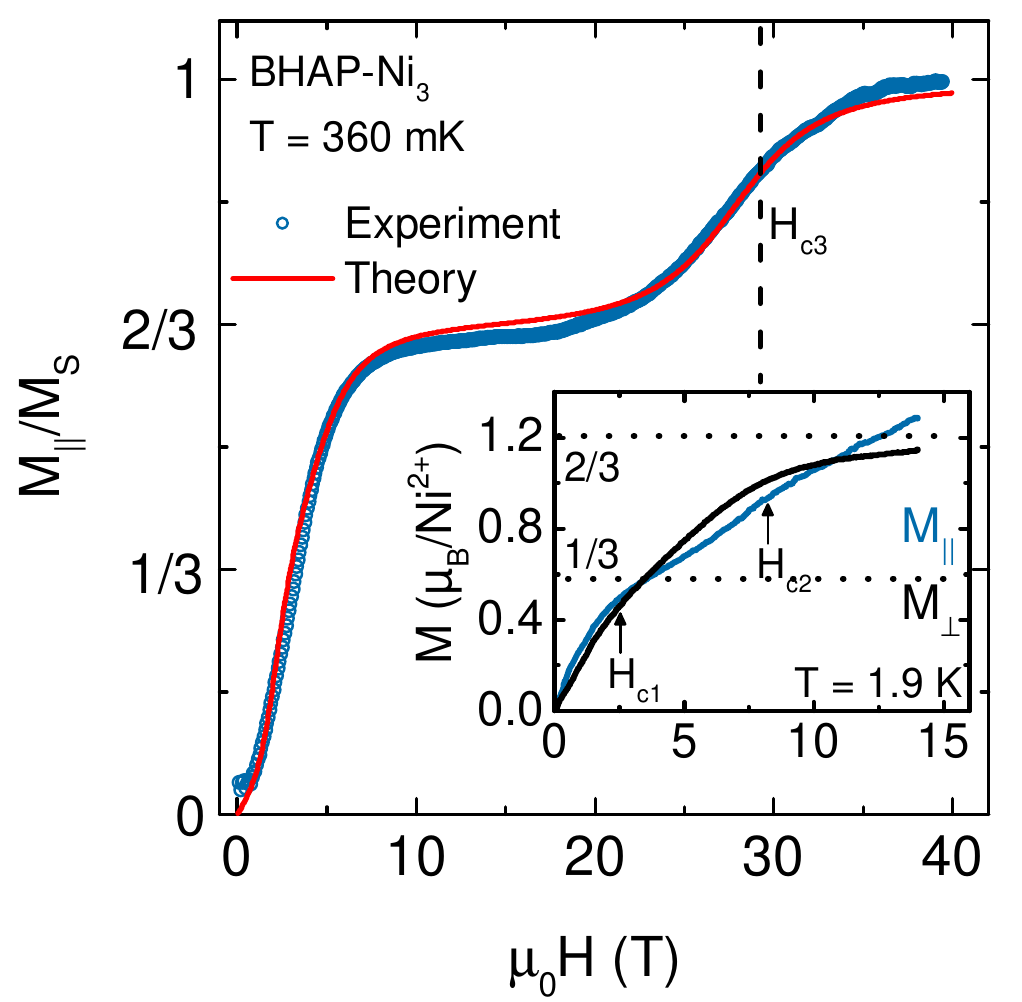}
	\caption{Isothermal field dependence of the magnetization ($M_{\parallel}$/$M_\mathrm{S}$) measured at $T$ = 360 mK in a pulsed magnetic field up to 40 T as well as the theoretical magnetization curve showing the 2/3-magnetization plateau. Inset: $M$($H$) curves obtained at 1.9 K using a vibrating sample magnetometer.}
	\label{Fig5}
\end{figure}

The inset of Fig. \ref{Fig5} shows the magnetic-field dependence of $M_\perp$ and $M_\parallel$ recorded at 1.9 K up to 14 T. Around $\mu_0H_{c1}$ = 2.5 T, $M_\parallel$ shows a weak anomaly which is almost invisible in $M_\perp$. With further increase in field, around $\mu_0H_{c2}$ = 8 T, $M_\perp$ enters to a plateau-like region whereas $M_\parallel$ continues to increase slowly. We would like to emphasize that this plateau above $H_{c2}$ does not correspond to the full saturation of the Ni moments as obtained from our density functional theory (DFT) calculations (1.75 $\mu_\mathrm{B}$/Ni$^{2+}$), rather it corresponds to 2/3 of the fully saturated moment. 
The very weak anomaly around $H_{c1}$ might be related to $1/3$ of full saturation. However, no clear plateau around this position could be seen.
For generating higher spin polarization, we performed $M$($H$) measurements up to 40 T using pulsed magnetic fields at 360 mK. Data were taken only with field applied within the plane of the triangles due to technical constraints. As shown in Fig. ~\ref{Fig5}, a 2/3 magnetization plateau is clearly visible between 7 and 20 T. The magnetization curve attains its full saturation above $\mu_0H_{c3}$ = 35 T where the spin triangle gets fully polarized with a $\vert1,1,1\rangle$ spin configuration. 
 
The absence of the weak anomaly near $H_{c1}$ in the pulsed-field data might be related to various factors related to the measurement technique.  
On the other hand, if the anomaly is stabilized via an order-by-disorder mechanism, it is expected to disappear with lowering of temperature \cite{Okuta2011,Smirnov2017}.
Pulsed field measurement was also performed at 2 K to check its agreement with the steady field measurement at 1.9 K (see Fig.~\ref{Fig:S2b} in the appendix). 
We would like to emphasize that the experimental observation of a 2/3 magnetization plateau in frustrated spin triangles is a novel phenomenon unlike the commonly observed 1/3 plateau. 
In a spin triangle, a 2/3 plateau can be for instance visible if the three spins form a $\vert1,1,0\rangle$ kind of arrangement. 

\par 
Next, we make a connection to our experimental results by carrying out a theoretical modeling of this interesting compound. As is evident from the density of states in GGA+U calculations~\cite{GGAU1,GGAU2,GGAU3} (see Appendix~\ref{App:DFT}), Ni is in its $d^8$ state, corresponding to a spin $S$ = 1. Due to the large distances between the Ni$_3$ centers, the three Ni spins form a nearly isolated triangle.
We describe the magnetic properties by a spin-model for these three Ni spins. 
Furthermore, the calculations suggest strongly anisotropic antiferromagnetic exchange interactions between the three Ni spins.

Based on these insights we formulate the following model:
\begin{eqnarray}
\label{Ham}
\begin{array}{r@{}l}
	\mathcal{H} &{}= 	   \sum_{\langle i,j\rangle} J_{ij} \mathbf{S}_i \cdot \mathbf{S}_j
				+ \sum_{\langle i,j\rangle} K_{ij} \left(\mathbf{S}_i \cdot \mathbf{S}_j\right)^2\\
			&{}	\ + \sum_{\langle i,j\rangle} D_{ij}\mathbf{\hat{d}}_{ij}\cdot \left(\mathbf{S}_i \times \mathbf{S}_j\right)
				+ 2\mu_B H \sum_i S^{z(x)}_i.
\end{array}
\end{eqnarray}
Here, $i$ denotes the three Ni spins and $\langle i,j\rangle$ the three Ni-Ni pairs of a triangle. The first term amounts to anisotropic Heisenberg exchange interactions between the Ni atoms with strengths $J_{ij}$. The effect of spin-orbit coupling has been taken into account through the Dzyaloshinskii-Moriya (DM) interaction in the third term of the Hamiltonian. The fourth term is related to the Zeeman energy while the second term denotes biquadratic spin interactions. The necessity of including the latter term will be discussed in the following section.

\par
Despite the seemingly simple structure of its metallic centre, the precise modeling of BHAP-Ni$_3$ is extremely difficult.
In the model presented here, we focus on the pronounced 2/3 plateau together with an almost vanishing 1/3 anomaly in the magnetization curve as main characteristic of BHAP-Ni$_{}3$. 
Although this model cannot account for all salient features at low magnetic field strengths, it correctly reproduces qualitatively the specific heat measured at zero and finite field strength. 
In Appendix~\ref{App:AltModel} we present an alternative model with two ferromagnetic and one antiferromagnetic exchange constants, which captures the low-field magnetization curve better, but whose specific heat curves are not compatible with our measurements (see Figs.~\ref{NewMod}).

\par
\begin{figure}[htbp]
	\includegraphics [width=\linewidth, angle=0]{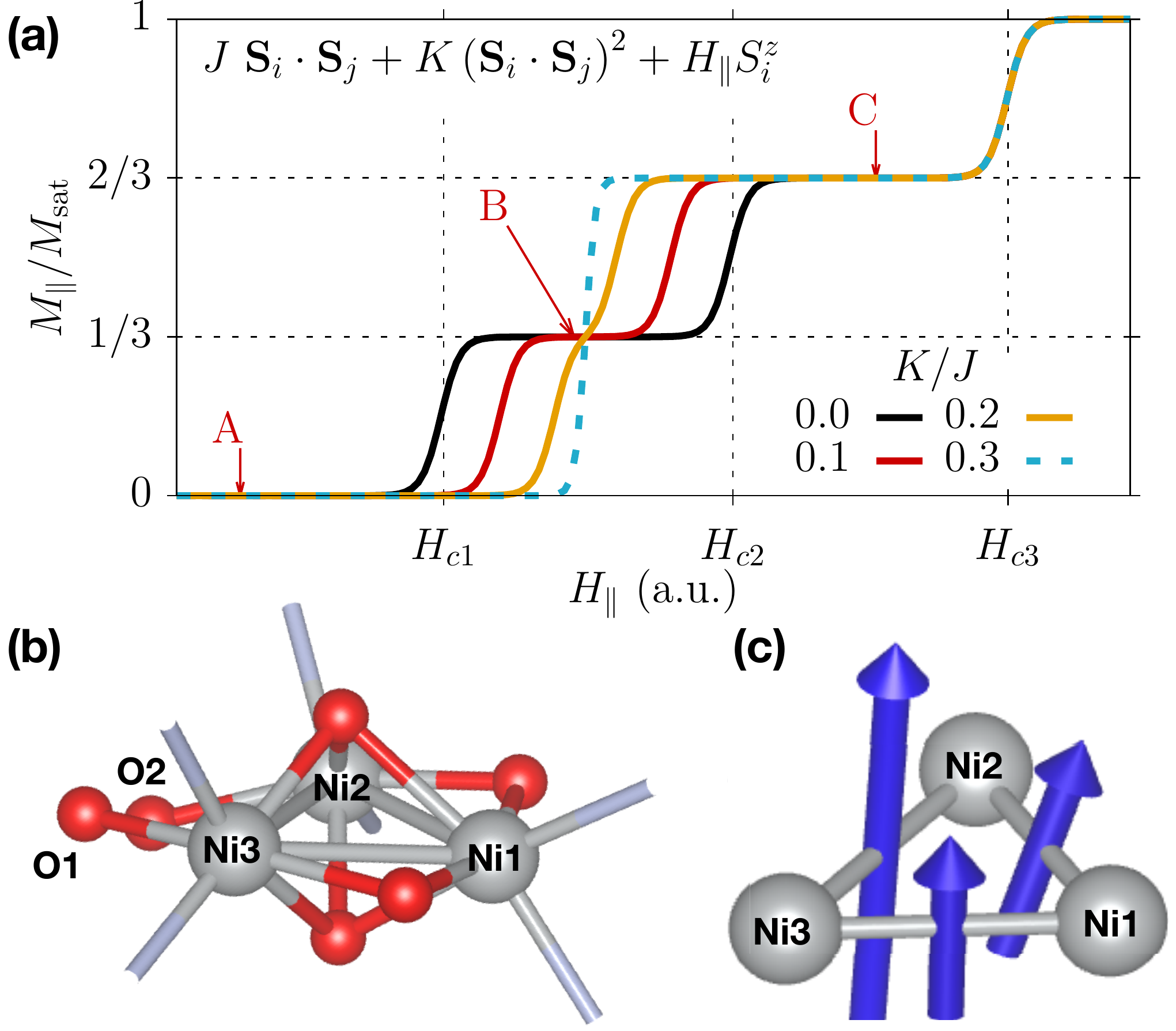}
	\caption{(a) Effect of a small biquadratic exchange term $K$ on the magnetization curve for the simplified model of a spin-1 triangle with isotropic spin exchange $J$. Ground-state compositions at the plateaus ($A,B,C$) are discussed in the text. (b) Illustration of the Ni triangle with two oxygen atoms O1 and O2 (red) not being shared between nickel atoms Ni1 and Ni3 (grey). Twisted ring exchange involving these four atoms might be responsible for the biquadratic interaction term in the spin model. (c) Sketch of the Dzyaloshinskii-Moriya vectors $\mathbf{d}_{ij}$ (blue) for the spin model.}
	\label{FigN1}
\end{figure}

 
The anisotropy of the antiferromagnetic exchange couplings itself is not sufficient to reproduce the characteristic plateau structures of the $M-H$ curve and even including a single-ion anisotropy term is not sufficient. 
However, recent numerical studies of the spin-1 bilinear-biquadratic Heisenberg (BBH) model on the isotropic triangular lattice revealed a 2/3 plateau in $M$ \cite{Lauchli06,Tsunetsugu06,Wessel15,Corboz18}. 
Motivated by these results, we include a biquadratic spin interaction of strength $K_{ij}$. Most interestingly, an AFM biquadratic term suppresses the 1/3 magnetization plateau also for our isolated Ni triangle and promotes the formation of a 2/3 plateau, see Fig.~\ref{FigN1}(a).
Here, however, the absence of strong inter-triangle interactions allows for the formation of a rather large $2/3$ plateau as compared to the aforementioned triangular lattice systems.

\par 
Biquadratic exchange terms occur naturally for systems with spin $S\geq1$ in fourth-order perturbation theory in the hopping \cite{Moriya60}, but usually they are much smaller than bilinear Heisenberg exchange terms. 
However, in some cases additional quasi-degenerate orbitals \cite{Mila2000} or twisted ring hopping processes \cite{Hotta2018} have been shown to be responsible for rather strong biquadratic exchange.
In the context of strongly frustrated antiferromagnets the effect of bond disorder and thermal and quantum fluctuations is sometimes described by a biquadratic term as well \cite{Zhitomirsky2002, Okuta2011, Maryasin2013, Smirnov2017}.
A derivation from a microscopic model to elucidate whether such mechanisms apply here or whether it is rather due to the integrating out of other degrees of freedom such as phonons is left for future work.

\par
As spin-orbit coupling is not negligible in Ni$^{2+}$ systems, we also include Dzyaloshinskii-Moriya (DM) terms of strength $D_{ij}$ between the Ni atoms. Since the Ni triangle is very asymmetric, the directions of the DM vectors $\mathbf{\hat{d}}_{ij}$ cannot be determined according to the  Moriya rules \cite{Moriya60}, but rather based on geometric arguments (see Appendix~\ref{App:DMI} for details). The directions of the DM vectors are indicated in Fig.~\ref{FigN1}(b). 
Finally, we add an in-plane (out-of-plane) magnetic field term to account for the applied magnetic field parallel (perpendicular) to the Ni-triangles.
More details on the model can be found in Appendix~\ref{App:Model}.

\par
Best agreement with experiment was obtained for exchange interactions $J_{31}=J_{12}\approx3.5$ K,\ $J_{23}\approx17.5$ K.
  ~Interestingly, we find that even small values of the biquadratic exchange $K$ already lead to a reduction of the size of a plateau at $M/M_\mathrm{S} = 1/3$. Large values of $K$ lead to a very steep jump between the $M/M_\mathrm{S} = 0$ and 2/3 region which sets bounds to the strength of the biquadratic exchange interaction. For good agreement with experiment, we have used a moderate biquadratic exchange of $K_{ij}\approx0.3J_{ij}$. 
The strength of the DM interaction was set to $D_{ij} \approx0.2J_{ij}$, which is of the same order of magnitude as 
$D_{ij} =0.12J_{ij}$ estimated via $\Delta g/g$ \cite{Moriya60} from our ESR measurement of $g$ {\footnote{Note also that a similar Cu$_3$ framework \cite{cho06} finds a DM interaction strength of the same order of magnitude.}}.

\par 
Without the DM term, the ground state of the system at small field strength $H<H_{c1}$ is given by the antisymmetric combination of the six permutations of the $\vert1,-1,0\rangle$ spin configuration.
For illustration purposes we sketch this situation in as A in Fig.~\ref{FigN1}(a) for the simplified model of an isotropic spin-1 triangle. 
Intermediate field strengths, $H_{c1}<H<H_{c2}$, stabilize a 1/3 plateau (B) with $\vert1,0,0\rangle$ and $\vert 1,1,-1\rangle$ 
configurations. 
The 2/3 plateau state (C) at field strength $H_{c2}<H<H_{c3}$ consists of the three permutations of the $\vert1,1,0\rangle$ spin configurations. 
However, due to the biquadratic Heisenberg term, the former 1/3 plateau is rather rudimentary and a clear attribution to the aforementioned state is not possible. 
Furthermore, since the DM term introduces spin-canting, contributions of other spin configurations get admixed and lead to more complicated states. 
Nevertheless, the 2/3 plateau still corresponds in good approximation to the $\vert1,1,0\rangle$ state with asymmetric contributions of the three spin permutations.

\begin{figure}[htbp]
\includegraphics [width=0.95\linewidth, angle=0]{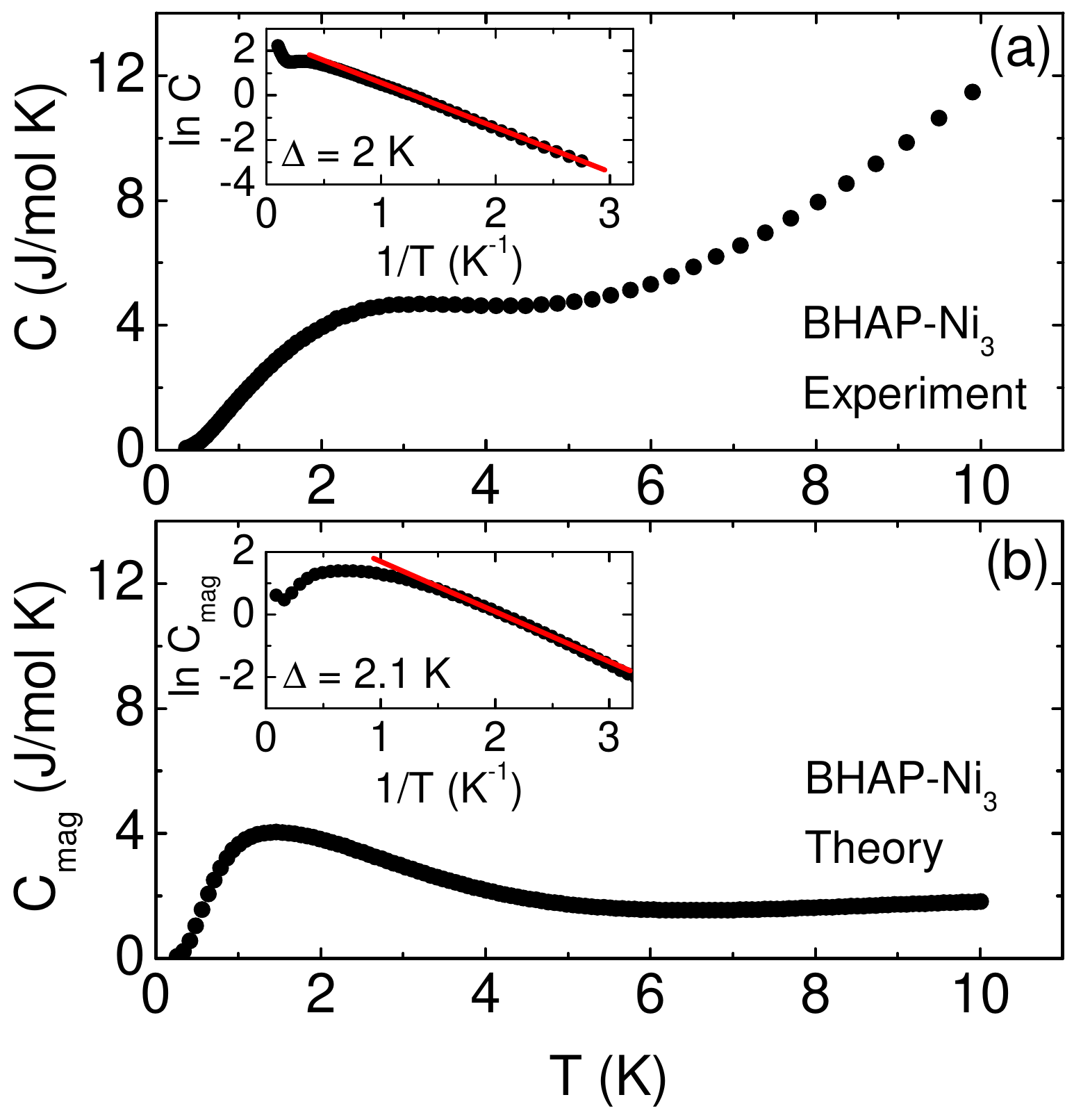}
\caption{(a) Temperature dependence of specific heat ($C$) as obtained experimentally. In the inset, $\ln C$ ($C$ in J/mol K) is plotted against $1/T$ down to 360 mK showing exponential behavior. The red line is the linear fit to the data which allows to extract the spin gap $\Delta$. (b) Magnetic contribution to the specific heat ($C_\mathrm{mag}$) calculated using the model with the parameters as described in the text (same as in Fig.~\ref{Fig5}). Inset shows $\ln C_\mathrm{mag}$ ($C_\mathrm{mag}$ in J/mol K) vs. 1/$T$ plot to show the similar exponential behavior with spin gap as of experiment.}
\label{Fig4}
\end{figure} 

\par
Figure \ref{Fig4}(a) shows the temperature dependence of the specific heat ($C$) measured down to 360 mK. A precise estimation of the phonon contribution to the $C$ is highly ambiguous, since in this case no non-magnetic isostructural compound is available. However, by implying a simple Debye-$T^3$ behavior to the data below 10 K, we found that the phonon contribution is negligibly small in the region of our interest (less than $\sim$1\% up to 2 K), making the spin contribution to be predominant. This behavior is validated by our theoretically calculated specific-heat (magnetic contribution). In zero field, the absence of any sharp anomaly signifying the absence of long-range magnetic order is in line with the observation from magnetometry. The $C$($T$) curve shows a broad hump-like feature around 2 K and an increase for temperatures above 5 K. Figure \ref{Fig4}(b) shows the theoretical calculation of the magnetic contribution to the specific heat ($C_\mathrm{mag}$) using the same parameter set utilized for calculating the magnetization (as in Fig.~\ref{Fig5}). It is in good agreement with the experimental $C$ up to 2 K. Above 2 K, where the phononic contributions to the specific heat are expected to become significant, the  theoretical $C_\mathrm{mag}$ deviates from the experimentally obtained total specific heat. In the inset of Fig. \ref{Fig4}(a), $\ln C$ is plotted against inverse temperature ($1/T$).  It is evident from the linear region of this plot that the specific heat shows exponential behavior below 1 K. Such a feature signifies the existence of a gap in the spin excitation spectra. Fitting the experimental data with $C \propto e^{-\Delta/k_\mathrm{B}T}$ reveals a spin gap of $\Delta$ = 2.0 K which nicely agrees with the theoretical value of 2.1 K as shown in the inset of Fig. \ref{Fig4}(b). The experimentally obtained $C$($T$) has good agreement with theoretical $C_\mathrm{mag}$ at low temperature even for measurements under applied magnetic field (Fig.~\ref{FigS5} in the Appendix). 
In contrast, the otherwise appealing alternative model described in Appendix~\ref{App:AltModel} cannot account for the low-temperature features of the specific heat.

\section{Summary and Conclusion}
In summary, we report combined experimental and theoretical studies on a new metal-organic system (BHAP-Ni$_3$) synthesized in single-crystalline form. This quantum magnet provides the unique opportunity to investigate the magnetism of a isolated and magnetically frustrated $S$ = 1 spin-triangle framework. High-field magnetometry reveals the existence of a robust 2/3 magnetization plateau. Extensive theoretical modeling suggests this exotic feature to be a consequence of the interplay between Heisenberg and biquadratic exchange interactions. Whereas the spin model can nicely reproduce the intriguing features of the magnetization curve, a microscopic derivation of the biquadratic spin terms is left for further investigations. The gapped nature of the magnetically disordered ground state has also been evidenced through specific-heat measurements in agreement with theory. In conclusion, usage of such molecular engineering could be extremely beneficial for the material science community to design and investigate novel quantum-magnetic frameworks of both fundamental and technical importances including spintronics and quantum computing. 

\section{Acknowledgment}
We acknowledge the support from HLD at HZDR, member of the European Magnetic Field Laboratory (EMFL), DFG through SFB 1143, and ZV 6/2-2. This work was also supported by a Consolidator Grant of the European Research Council (Project CorrelMat-617196) and  supercomputing time at IDRIS/GENCI Orsay (Project No.  t2018091393). We are grateful to the CPHT computer support team. S.K. thanks the Department of Science and Technology, Govt. of  India for providing funding through INSPIRE project (DST/INSPIRE/04/2016/000431) and IUAC, New Delhi, India for providing computational facilities. Sushila thanks CSIR-INDIA for fellowship. R.P. thanks DST-India for INSPIRE Project.


\appendix
\section{Overview}
We would like to provide more technical  details  of  our  work  on  [2-[Bis(2-hydroxybenzyl)aminomethyl]pyridine]Ni(II)-trimer (BHAP-Ni$_3$) concerning both its experimental and theoretical aspects. First, we discuss the sample preparation and characterization. Then, we present details on the density functional theory (DFT) calculations and the effective model, which has allowed us to calculate the theoretical magnetization and specific-heat curves presented in the main paper. 
Finally, we present an alternative model that can better account for the low-field magnetization and susceptibility curves of BHAP-Ni$_3$, but fails to reproduce salient features of the specific heat curve. 

\section{Synthesis procedure}
\label{App:Sample}
\par
All chemicals were used as obtained without further purification. UV-vis spectra were recorded on a Perkin Elmer UV-Vis-NIR spectrometer. 
Infrared spectra were recorded using a Perkin-Elmer Spectrum Two FT/IR spectrometer. The ligand  2-[Bis(2-hydroxybenzyl)aminomethyl]pyridine (BHAP) was synthesized according to the literature procedure~\cite{lig}.
\par
The ligand solution was prepared by stirring 2-[bis(2-hydroxybenzyl]pyridine (0.032 g, 0.1 mmol) in 10 ml dichloromethane (DCM) in a round bottom flask, and 2-3 drops of triethylamine were added. The mixture was stirred for 15 minutes, then methanolic solution of Nickel nitrate hexahydrate (0.029 g, 0.1 mmol) was added to the ligand solution and stirred for 12 hours. Then, the solvent was evaporated completely with the help of a rotary evaporator and a green colored solid was obtained. That green solid was dissolved in acetone and layered with methanol in a crystallization tube for slow evaporation. Tiny green crystals of typical size $\sim1 \times 0.4 \times 0.2$ mm$^3$  were obtained (see Fig.~S1) after 10-12 days. The yield was 0.022 g (60\%). FT-IR results are as follows (KBr, cm$^{-1}$): 3068, 1679, 1408, 1252, 1032, 802, 741, 660, 571, and 519. The UV-Vis in DCM (10$^{-6}$ M)  shows absorptions [using $\lambda_{\mathrm{max/nm}}$(log$_\mathrm{e}$) format] at 418(3.25) and 625(1.24). In the IR study, clear signatures of the Ni-O stretching frequency were observed.

\begin{figure}[]
	\includegraphics [width=0.9\linewidth, angle=0]{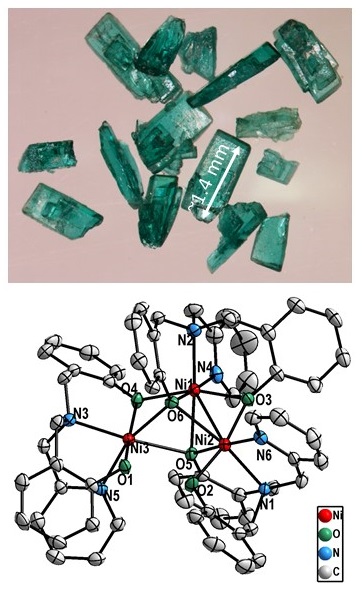}
	\caption{\footnotesize Top panel: Plate like tiny BHAP-Ni$_3$ single crystals. Bottom panel: Perspective view of BHAP-Ni$_3$ crystal structure obtained by solving the single-crystal x-ray diffraction data, showing 50\% thermal ellipsoids for all non-hydrogen atoms at 293 K. Solvents are omitted for clarity.}
	\label{FigS1}
\end{figure}  

\section{Single-crystal x-ray diffraction}
\label{App:XRD}
\par
Single-crystal x-ray diffraction data were collected with a SuperNova Diffractometer equipped with a HyPix3000 detector from Rigaku Oxford Diffraction. Data collection and reduction were performed with the in-built program suite (CrysAlisPro 1.171.39.33c (Rigaku OD, 2013)) and an absorption correction (multi-scan method) was also done. The crystal structure was solved by the direct method using SHELXS-97 and was refined on $F^2$ by the full-matrix least-squares technique using the SHELXL-2018/3~\cite{shelx1,shelx2} program package on the WINGX~\cite{wingx} platform. 
All non-hydrogen atoms were refined anisotropically. Hydrogen atoms were fixed at their stereo-chemical positions and were riding with their respective non-hydrogen atoms with SHELXL default parameters. 
Detailed information can be found in Tables I, II, III, and in the supplementary material~{\footnote{Crystallographic Information File (CIF) has been provided as supplementary material for structural details}}.

 \begin{figure}[]
 	\includegraphics [width=0.9\linewidth, angle=0]{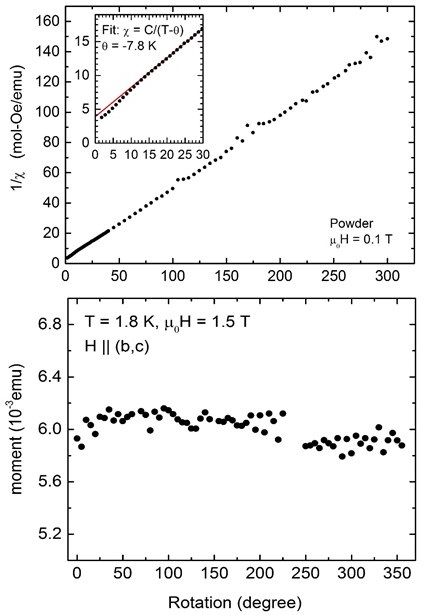}
 	\caption{\label{Fig:S2}\footnotesize Top panel: Temperature variation of inverse dc magnetic susceptibility (1/$\chi$ = $H$/$M$) for powder sample. Inset shows Curie-Weiss fit to the powder data. Bottom panel: Angular dependence of magnetic moment ($m$) in the ($b$,$c$) plane measured at 1.8 K using single crystal.}
 	\label{FigS2}
 \end{figure}  


\begin{figure}[]
	\includegraphics [width=0.9\linewidth, angle=0]{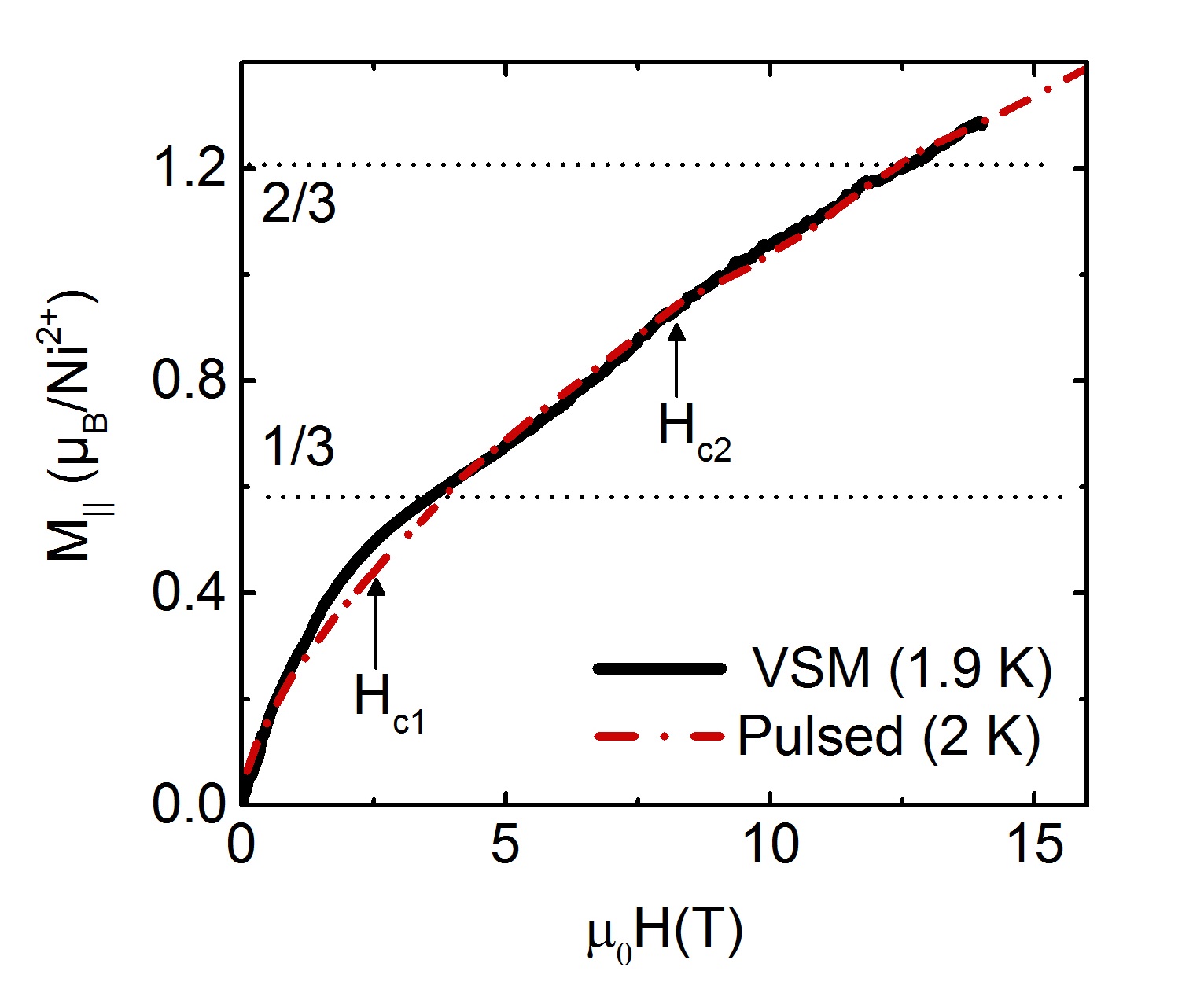}
	\caption{\label{Fig:S2b}\footnotesize Magnetization as a function of applied magnetic field for both the VSM and the pulsed field measurements at a temperature of $\sim$2 K.}
	\label{FigS2}
\end{figure}  


\begin{table}[t]
\caption{\small{Crystal data and data collection parameters for BHAP-Ni$_3$.}}
\label{table1}
\begin{tabular} {cc}
\hline\\
CCDC No.	&1870984 \\     
$T$(K)  	&293 \\
Formula	&C$_{63}$H$_{70}$N$_6$Ni$_3$O$_{12}$ \\
Formula weight	&1279.32 \\
Color and Habit	&Green \\
Crystal system	&Monoclinic \\
Space group	&$P2(1)/n$ \\
$a$(\AA) 	&12.7361(2) \\
$b$(\AA)	&29.4086(4) \\
$c$(\AA)	&15.3701(2) \\
$\alpha$($^\circ$)		&90 \\
$\beta$($^\circ$)	&96.824(2) \\
$\gamma$($^\circ$)	&90 \\
$V$(\AA$^3$) 	&5716.10(14) \\
Radiation: $\lambda$, \AA 	&Mo $K\alpha$, 0.71073 \\
$Z$	&4 \\
$d$(g-cm$^{-3}$)	&1.487 \\
$\mu$(mm$^{-1}$)	&1.049 \\
$F$(000)	&2680 \\
No. of unique data 	&10064 \\
No. of restraints	&0 \\
No. of parameters refined	&745 \\
GOF on $F^2$	&1.073 \\
$R1a$ [$I$$>$2$\sigma$($I$)]	&0.0461 \\
$R1a$ (all data)	&0.0535 \\
$wR2b$ (all data)	&0.1566 \\\\
\hline\\
\end{tabular}
\end{table}
\begin{table}[t]
\caption{\small{List of important bond distances (\AA) in BHAP-Ni$_3$.}}
\label{table2}
\begin{tabular} {cccc}
\hline\\
Ni(1)-O(3) 	&2.010(2)		 &Ni(2)-O(3) 	&2.048(2) \\
Ni(1)-O(4) 	&2.046(2)		 &Ni(2)-N(6) 	&2.055(3) \\
Ni(1)-N(2) 	&2.048(3)		 &Ni(2)-N(1) 	&2.072(3) \\
Ni(1)-N(4) 	&2.055(3)		 &Ni(2)-O(6) 	&2.298(2) \\
Ni(1)-O(6) 	&2.058(2)		 &Ni(3)-O(4) 	&1.987(2) \\
Ni(1)-O(5) 	&2.177(2)		 &Ni(3)-O(1) 	&2.037(2) \\
Ni(1)-Ni(2) &2.8623(5)	 & Ni(3)-N(5) 	&2.060(3) \\
Ni(1)-Ni(3) &2.9097(5)	 & Ni(3)-N(3) 	&2.137(3) \\
Ni(2)-Ni(3) &3.411(5)          &Ni(3)-O(6) 	&2.142(2) \\
Ni(2)-O(2) 	&2.025(2)		 &Ni(3)-O(5) 	&2.263(2) \\
Ni(2)-O(5) 	&2.043(2)		 &                     &              \\\\
\hline\\
\end{tabular}
\end{table}
\begin{table}[t]
\caption{\small{List of selected bond angles (in degree) in BHAP-Ni$_3$.}}
\label{table3}
\begin{tabular} {cccc}
\hline\\
O(3)-Ni(1)-O(4)	&159.62(9)	 &O(5)-Ni(2)-N(1)	      &93.85(10) \\
O(3)-Ni(1)-N(2)	&95.59(10)	 &O(3)-Ni(2)-N(1)	      &100.07(10) \\
O(4)-Ni(1)-N(2)	&101.93(10)	&N(6)-Ni(2)-N(1)	      &83.18(11) \\
O(3)-Ni(1)-N(4)	&94.36(10)	&O(2)-Ni(2)-O(6) 	 &89.33(8) \\
O(4)-Ni(1)-N(4)	&97.90(10)	&O(5)-Ni(2)-O(6)	      &68.83(8) \\
N(2)-Ni(1)-N(4)	&83.30(11)	&O(3)-Ni(2)-O(6)	      &79.29(8) \\
O(3)-Ni(1)-O(6)	&86.21(9)		&N(6)-Ni(2)-O(6)	      &114.14(9) \\
O(4)-Ni(1)-O(6)	&82.40(9)		&N(1)-Ni(2)-O(6)	      &162.66(9) \\
N(2)-Ni(1)-O(6)	&93.66(10)     &O(2)-Ni(2)-Ni(1)	 &124.60(7) \\
N(4)-Ni(1)-O(6)	&176.94(10)	&O(5)-Ni(2)-Ni(1)	       &49.28(6) \\
O(3)-Ni(1)-O(5)	&80.85(9)		&O(3)-Ni(2)-Ni(1)	      &44.60(6)\\
O(4)-Ni(1)-O(5)	&79.49(8)		&N(6)-Ni(2)-Ni(1) 	 &132.43(8) \\
N(2)-Ni(1)-O(5)	&164.44(9)	&N(1)-Ni(2)-Ni(1)	      &123.45(8)\\
N(4)-Ni(1)-O(5)	&111.99(10)	&O(6)-Ni(2)-Ni(1)      &45.38(5) \\
O(6)-Ni(1)-O(5)	&71.06(8)		&O(4)-Ni(3)-O(1)	      &164.84(9) \\
O(3)-Ni(1)-Ni(2)	&45.69(6)		&O(4)-Ni(3)-N(5)	      &109.75(10) \\
O(4)-Ni(1)-Ni(2)	&114.68(6)	&O(1)-Ni(3)-N(5)	      &85.16(10) \\
N(2)-Ni(1)-Ni(2)	&122.73(8)	&O(4)-Ni(3)-N(3)	      &90.99(10) \\
N(4)-Ni(1)-Ni(2)	&129.48(8)	&O(1)-Ni(3)-N(3)	      &88.49(10) \\
O(5)-Ni(1)-Ni(2)	&45.36(6)		&N(5)-Ni(3)-N(3)	      &81.31(10) \\
O(3)-Ni(1)-Ni(3)	&118.11(7)	&O(4)-Ni(3)-O(6)	      &81.68(9)  \\
O(4)-Ni(1)-Ni(3) &43.04(6)        &O(1)-Ni(3)-O(6)	      &84.71(8) \\
N(2)-Ni(1)-Ni(3) &120.97(8)      &N(5)-Ni(3)-O(6)	      &161.20(10) \\
N(4)-Ni(1)-Ni(3)	&134.33(8)    &N(3)-Ni(3)-O(6)	      &114.18(9) \\
O(6)-Ni(1)-Ni(3) &47.35(6)       &O(4)-Ni(3)-O(5)	      &78.63(8) \\
O(5)-Ni(1)-Ni(3) &50.35(6)      & O(1)-Ni(3)-O(5)	      &102.40(8) \\
Ni(2)-Ni(1)-Ni(3)	&72.438(14) &N(5)-Ni(3)-O(5)	      &98.98(9) \\
O(6)-Ni(1)-Ni(2)	&52.65(6)	   &N(3)-Ni(3)-O(5)	      &169.10(9) \\
O(2)-Ni(2)-O(5) &90.64(9)      &O(6)-Ni(3)-O(5)	      &67.92(8) \\
O(2)-Ni(2)-O(3)	&168.40(9)  &O(4)-Ni(3)-Ni(1)	      &44.62(6) \\
O(5)-Ni(2)-O(3)	&83.24(9)    &O(1)-Ni(3)-Ni(1)	      &125.74(6) \\
O(2)-Ni(2)-N(6)  &89.68(10)   &N(5)-Ni(3)-Ni(1)	      &135.09(8) \\
O(5)-Ni(2)-N(6) &177.01(10)  &N(3)-Ni(3)-Ni(1)	      &125.41(7) \\
O(3)-Ni(2)-N(6) &96.94(10)    &O(6)-Ni(3)-Ni(1)	      &44.96(6) \\
O(2)-Ni(2)-N(1) &90.13(10)    &O(5)-Ni(3)-Ni(1)	      &47.78(6) \\
\hline\\
\end{tabular}
\end{table}
\section{Magnetometry and Electron Spin Resonance}
\label{App:DCM}

DC magnetization measurements were performed with both powder sample and using oriented single crystals. 
These measurements were performed using either VSM or a SQUID magnetometer. 
The top panel of Fig.~\ref{Fig:S2} shows the temperature variation of the inverse dc magnetic susceptibility (1/$\chi$ = $H$/$M$) as obtained using a powder sample. 
The bottom panel shows the absence of any angular dependence of the magnetic moment ($m$) in the ($b$,$c$) plane. 

In Fig.~\ref{Fig:S2b} we compare the magnetization as a function of applied magnetic field for the VSM and the pulsed field measurements at $\sim$2 K. The good agreement between the two techniques demonstrates the validity of data taken with these complimentary approaches.

\par

Electron spin resonance measurements were performed using a high-field transmission-type ESR spectrometer, similar to that described in Ref.\cite{esr}. A set of VDI microwave sources (Virginia Diodes, Inc.) was used. The measurements  were done in the Faraday configuration with magnetic field applied in the plane of the Ni triangles. The spectra were recorded in the 60-150 GHz  frequency range  at a temperature of 80 K.

\par

\section{Density-Functional-Theory}
\label{App:DFT}

\par
The density-functional-theory (DFT) calculations were performed using the plane-wave basis set as implemented in the pseudopotential framework of the Vienna ab initio simulation package (VASP) \cite{Kresse93, Kresse96}. We employed the generalized gradient approximation (GGA) exchange-correlation functional following the Perdew-Burke-Ernzerhof prescription \cite{PBE96}. For the GGA+U calculations, we followed the formulation described in Refs. \cite{GGAU1,GGAU2,GGAU3}. For the plane-wave basis, a 600 eV plane-wave cutoff was applied. A $k$-point mesh of 4$\times$4$\times$4 in the Brillouin zone was used for self-consistent calculations. In the calculations, the spin-orbit coupling term was included in the scalar relativistic form as a perturbation to the original Hamiltonian. 

\par
Using GGA+U ($U_\mathrm{eff}$ = 5 eV at the Ni site), we calculated the spin-polarized atom projected electronic density of states (DOS) as shown in Fig. \ref{FigS2}. From the DOS, it is evident that the Ni-3$d$ states are dominating near the Fermi energy with a very strong hybridization with the O-2$p$ states. Due to the highly distorted nature of the mixed ligand octahedra, Ni-$d$ states are strongly mixed up. The Ni-$d$ states are completely filled in the majority spin channel and partially in the minority spin channel. The Ni-$d$ integrated DOS clearly shows that the majority and the minority spin channels are filled with five and three electronic states, respectively. This is consistent with the Ni atoms being in their nominal 2+ valence state ($d^8$) with $S$ = 1. It is to be noted that the spin-polarized calculations within the GGA, assuming small Coulomb correlation ($U$) at the 3$d$-Ni sites, drive the band structure with a gap of $\sim$1 eV. 

\par
To probe the non-collinear contribution we calculated the effect of spin-orbit coupling (SOC) through GGA+U+SOC for the different spin quantization axes. From our calculations, we found that the Ni site has considerable orbital contribution with a magnetic moment of 0.08 $\mu_\mathrm{B}$/site parallel to the spin moment 1.67 $\mu_\mathrm{B}$/site. This substantial orbital moment is unexpected for the Ni$^{2+}$ ($d^8$) configuration in the octahedral environment due to a completely quenched orbital degree of freedom, although a similar orbital moment of Ni-$d^8$ has already been reported previously\cite{jia07,sar10}. Therefore, the presence of finite and substantial contribution of orbital moment needs to be justified as an induced mechanism via ligands of O-$2p$ states. Magnetocrystalline anisotropy energies are obtained by analyzing the energy difference of the spin quantization axes between [100] and [011]. The obtained anisotropy energy is very small ($\sim$-0.06 meV/Ni) and favors [011] easy-plane type single-ion anisotropy. Since its size is negligible compared to the smallest exchange interaction used in our model Hamiltonian, a single-ion anisotropy term has not been considered in the modeling. We have cross-checked the DFT+U results by doing additional calculations with $U_\mathrm{eff}$ values between 2 eV and 5 eV at the Ni$^{2+}$ site. In contrast to the expected changes in the magnitudes of the $J$ values, both the sign and the trend of $J$ values remained unaltered with the variation of $U$. We also found that the changes in the saturation magnetic moment per Ni$^{2+}$ site is less than  10\%, keeping the total magnetic moment per unit cell intact.

\section{Models and exact diagonalization results}
\label{App:Model}
\par
Based on the outcome of our DFT calculations, we model the central Ni$_3$ unit by a triangle of spins with $S = 1$. 
The effective spin model is then solved at zero and finite temperatures using exact diagonalization (ED). 
Quantities like magnetization $M$, susceptibility $\chi$, and specific-heat $C_{\mathrm{mag}}$ have been calculated and compared with experiment. 
Together with insights gained from the DFT calculations, this comparison was used to refine the model and determine estimates for the model parameters. 
As this procedure in principle leads to a large set of parameters, the model and the number of parameters were reduced to a minimal set necessary to qualitatively reproduce the experiments. 

In the following, we derive and motivate the different terms of the model and discuss the values of the model parameters.
Finally, we will also present an alternative model, which is even more simplistic and still capable of capture part of the magnetic properties of BHAP-Ni${}_3$, but less precise in the description of the specific heat data.

\subsection{Linear and Quadratic Heisenberg Terms}
The triangular unit consists of three Ni atoms with edge-sharing O-N-octahedra around them. Ni1-Ni2 and Ni1-Ni3 share three oxygen atoms, which mediate an effective superexchange between the Ni sites. Since the Ni-O-Ni bonding angles are close to $90^{\circ}$, they lead to an overall very small antiferromagnetic exchange and also have to be taken into account for additional asymmetric exchange contributions (see next section). 
In case of the Ni2-Ni3 bond, only two oxygen ligands are shared and form an angle of $129^{\circ}$, see Fig.~\ref{FigS3}. Here, the bonding angles are even larger, which should lead to a substantially larger antiferromagnetic exchange. This is why we model the exchange part of the Ni-triangle via
\begin{equation}
	\sum_{\{i,j\}} J_{ij}\ \mathbf{S_i}\cdot\mathbf{S_j},
\end{equation} 
where $\{i,j\}$ runs over the three Ni-Ni bonds. Indeed, the best agreement of our model calculations with experiment have been obtained for two small antiferromagnetic exchange interactions $J_{31}=J_{12}\approx$ 3.5 K and a substantially larger interaction strength $J_{23}\approx$ 17.5 K.
In the corresponding magnetization curve it is mainly $J_{23}$ which sets the critical magnetic-field value for the transition from the 2/3 plateau to full magnetization, whereas the value of $J_{31}$ and $J_{12}$ influence the position of the transition to the 1/3 anomaly.

\par
In order to reproduce a strong 2/3 magnetization plateau our model has to be augmented by a biquadratic term. This term is known to cause a 2/3 plateau in the bilinear-biquadratic Heisenberg model on a triangular lattice \cite{lac11,chu91,zhi00,hen89,kam18} in the limit of strong biquadratic interactions. For $S\geq1$, it occurs naturally in fourth-order perturbation theory in the hopping $t_{ij}$ and is, therefore, usually suppressed as compared to second-order processes such as ordinary exchange interactions by $\sim(t/U)^2$ \cite{Moriya60}. In some cases, however, the intricate interplay of multi-orbital hopping processes, Hubbard interaction strength $U$ and Hund's interactions $J$ can lead to a relatively large biquadratic term \cite{Mila2000}. Note, however, that the interaction derived by Mila and Zhang is ferromagnetic. AFM biquadratic terms can be obtained from twisted ring exchange processes as recently shown by Tanaka et al. \cite{Hotta2018}. A rigorous derivation of the term for BHAP-Ni${}_3$ from a microscopic model is left for future studies. In our model, however, we included a biquadratic term with an interaction strength $K_{ij}$ proportional to the antiferromagnetic exchange strengths $J_{ij}$:

\begin{equation}
	\sum_{\{i,j\}}K_{ij}\left(\mathbf{S_i}\cdot\mathbf{S_j}\right)^2.
\end{equation}

In the spin-1 triangle, the biquadratic term stabilizes a distinct 2/3 plateau for moderate values of biquadratic interactions already ($K_{ij}/J_{ij}\sim0.3$). At the same time it also reduces the size of the 1/3 plateau, as shown in Fig. 3 of the main text.

\par

\begin{figure}[t]
	\includegraphics [width=0.9\linewidth, angle=0]{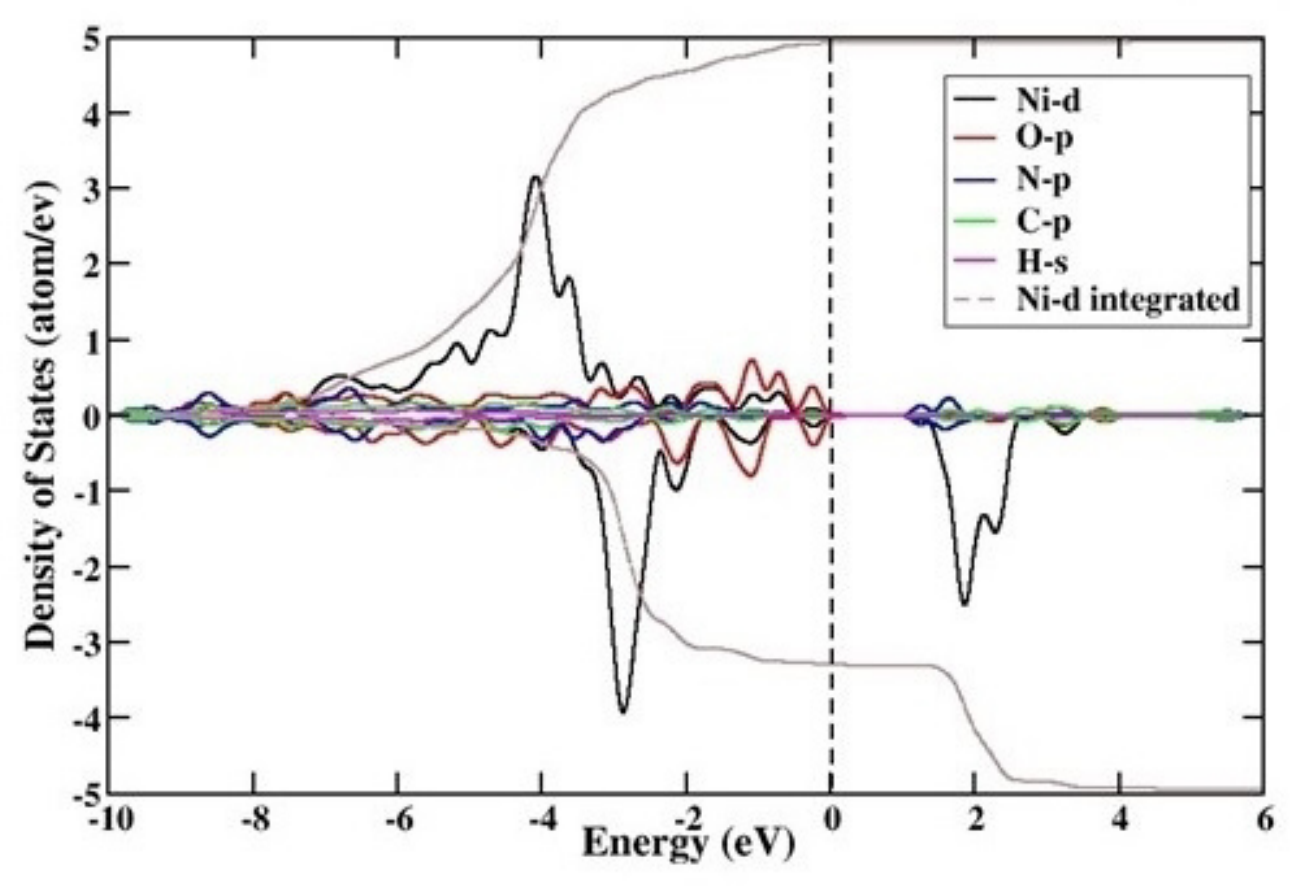}
	\caption{\footnotesize Calculated GGA+U ($U_\mathrm{eff}$ = 5 eV) spin polarized density of states (DOS) projected onto relevant orbital of different atoms in the unit cell are shown. The Fermi energy was set at the zero in the energy scale. The Ni-$d$ integrated DOS suggest that below the Fermi energy all the five $d$-states in majority spin channel and three $d$-state in minority are filled, while remaining two states in minority $d$-states are empty.}
	\label{FigS2}
\end{figure} 

\subsection{Dzyaloshinskii-Moriya Interaction}
\label{App:DMI}

\begin{figure}[tb]
	\includegraphics[width=.99\linewidth]{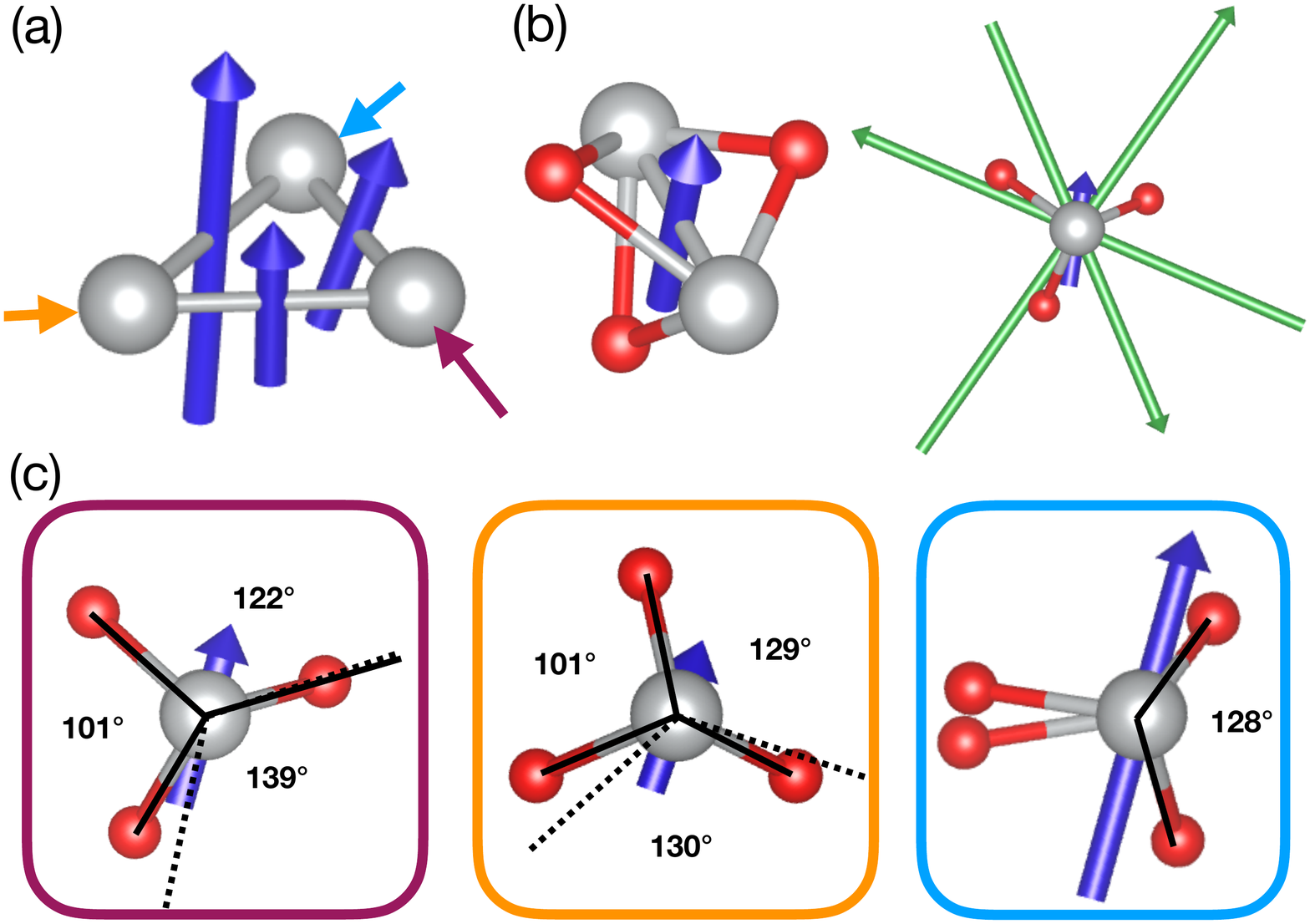}
	\caption{\footnotesize Construction of the DM vectors (a) of the Ni-triangle (grey spheres). (b) Two Ni atoms are connected via three oxygen atoms (red spheres), which mediate the superexchange and are considered for the DM interactions. Each Ni-O-Ni triangle contributes a DM component (green arrows) perpendicular to it resulting in the DM vector (blue). (c) View along the Ni-Ni axes indicated in (a): The three O atoms in the left and center panel deviate from a three-fold symmetric arrangement (dashed lines) and lead to a finite net DM vector. In the right panel, only two O atoms are shared between the Ni atoms.}
	\label{FigS3}
\end{figure}

The spin-orbit coupling constant of nickel is $\zeta_\mathrm{SOC}$(Ni)$\sim$80meV \cite{Cole70}, which is of comparable size as the one of copper, $\zeta_\mathrm{SOC}$(Cu)$\sim$100meV \cite{VJ96}, where spin-orbit effects were found to be important in a Cu$_3$ framework \cite{cho06}.
Due to the considerable spin-orbit coupling of nickel and because the oxygen octahedra are asymmetrically distorted, the emerging effective spin model is expected to contain non-zero Dzyaloshinskii-Moriya (DM) interactions of the form

\begin{equation}
	\sum_{\{i,j\}} D_{ij} \mathbf{\hat{d}_{ij}} \cdot (\mathbf{S_i}\times\mathbf{S_j}),
\end{equation}

where $D_{ij}$ is the DM interaction strength and $\mathbf{\hat{d}_{ij}}$ the normalized DM vector between the Ni atoms $i$ and $j$. In a crystal, one uses symmetry arguments to determine the direction of the DM vectors, the so-called Moriya rules \cite{Moriya60}. Here, however, we are dealing with a isolated trimer unit where the ligand fields are very asymmetric, see Fig.~\ref{FigS3}. Since the arrangement of the shared O atoms around the Ni3-Ni1 and Ni1-Ni2 axes deviates from a three-fold rotation-invariant configuration, the individual components of each Ni-O-Ni triangle do not compensate each other fully and lead to a finite net DM vector. For the Ni2-Ni3 bond, the two contributions to the DM vector corresponding to the Ni-O-Ni triangles lead to a quite large effective DM vector as compared to the other two Ni-Ni bonds.

\par
As shown in Fig.~\ref{FigS3}(a), all three DM vectors mainly have a component perpendicular to the Ni triangle. To simplify the model calculation, we used these directions of the DM vectors and set the DM interaction strengths proportional to the exchange strengths $J_{ij}$. Since a rigorous determination of the DM vectors would require a microscopic orbital-dependent treatment of the hopping processes leading to the effective exchange interaction, the vectors shown in Fig.~\ref{FigS3} are only an estimate. A more precise derivation is beyond the scope of the present article and left for future work.

\section{Specific-heat: Experiment and Modeling}
\label{SH}

\begin{figure}[]
	\includegraphics [width=0.9\linewidth, angle=0]{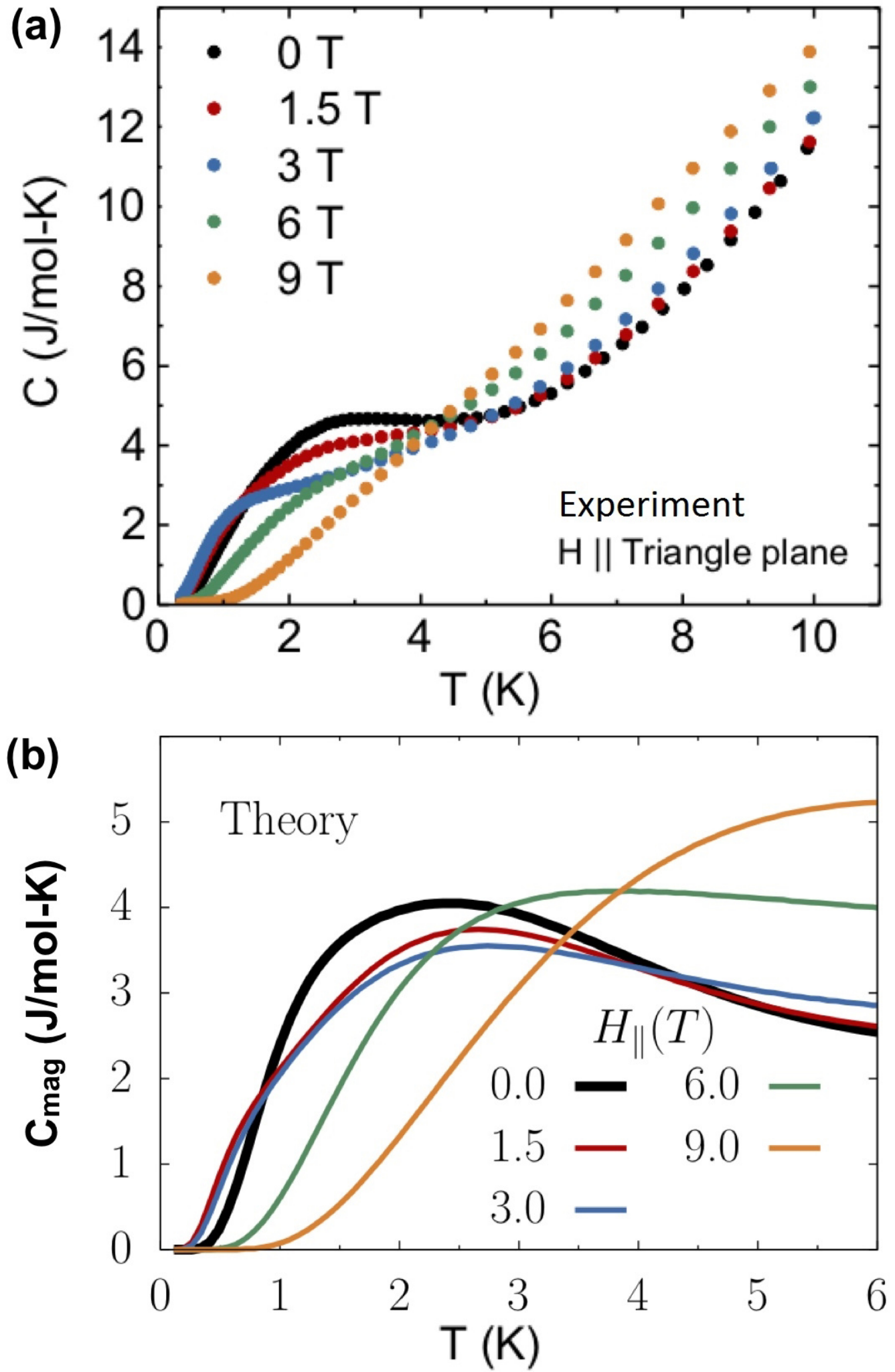}
	\caption{\footnotesize (a) Experimentally obtained specfic heat ($C$) data plotted as a function of temperature. Measurements were performed both in zero field and with different field up to 9 T. (b) Theoretically obtained magnetic contributions of the specific-heat ($C_\mathrm{mag}$) both in zero field and in presence of magnetic field.}
	\label{FigS5}
\end{figure}  
Specific-heat ($C$) measurements were performed down to $360$ mK both in zero field condition and in presence of an external magnetic field applied in the plane of triangle. 
The top panel of Fig.~\ref{FigS5} shows experimental specific-heat data as a function of temperature. 
The bottom panel is the simulated magnetic contribution to the specific-heat under similar protocols as of the experiment ($C_\mathrm{mag}$) using the spin Hamiltonian and the same set of parameters (described in the main text, equation 1) used to fit the $M$($H$) curve. 

In the low-temperature region where the estimated lattice contribution to the specific heat is negligible (less than $1\%$ for $T<2$K using a Debye-$T^3$ model) the behavior of the specific heat is qualitatively captured by our model calculation of the magnetic contribution $C_{\mathrm{mag}}$.
At higher temperatures, the specific heat is dominated by the lattice contribution, which leads to an increase of $C$, whereas the calculated $C_{\mathrm{mag}}$ decreases.

\section{Low-Field Properties of the Model}
\label{LF}
In addition to the quantities presented in the main text, we provide here additional information on the magnetic properties of the model at small magnetic field strengths $\mu_0H$.
Whereas the overall agreement of the magnetic and thermal properties of the model and experiment are good, it does not capture some salient features of the magnetic properties at very small magnetic field strengths correctly.

\begin{figure}[]
	\includegraphics [width=0.9\linewidth, angle=0]{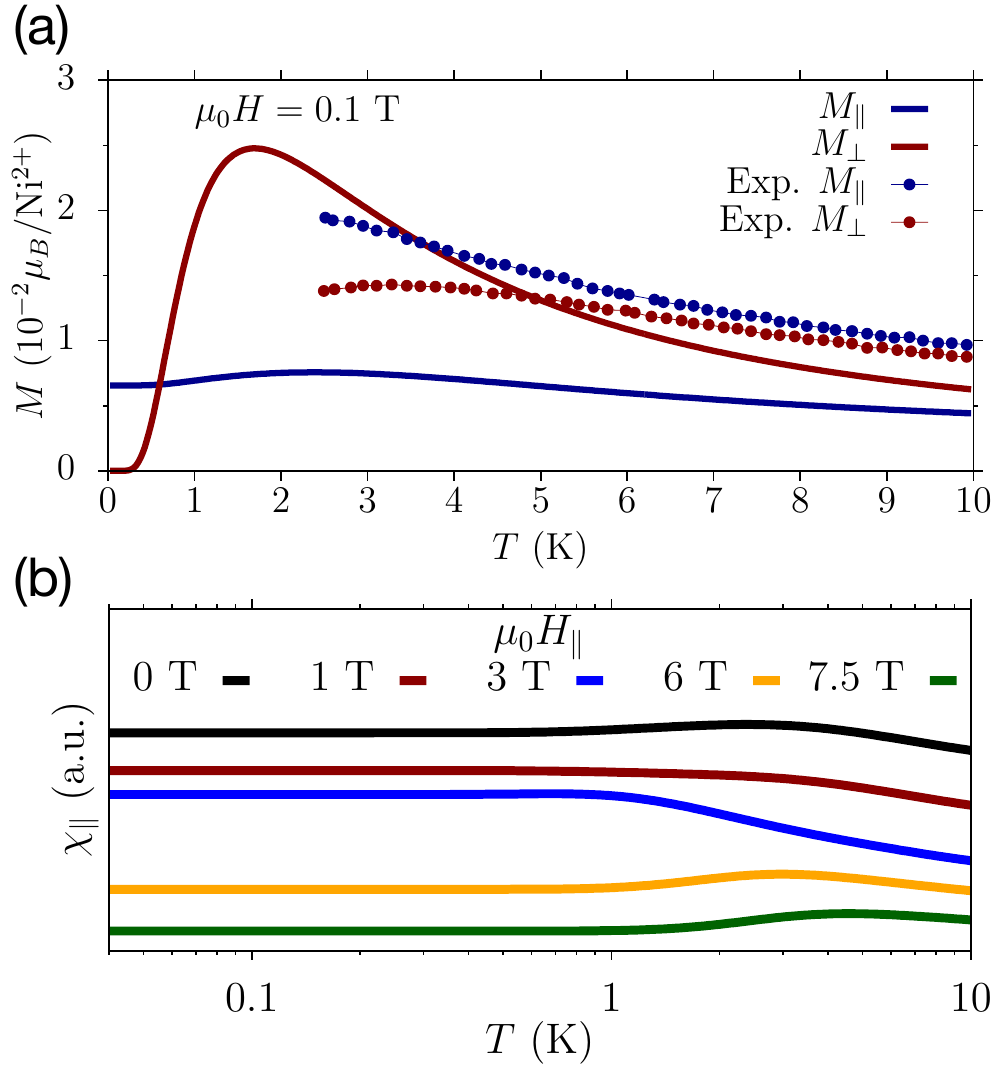}
	\caption{\footnotesize Temperature dependence of the magnetic properties of the model. (a) Calculated magnetization as a function of temperature for an external magnetic field strength of $\mu_0H=0.1$T. (b) In-plane susceptibility as a function of temperature for different magnetic field strengths. The curves have been shifted for better visibility.}
	\label{FigA1}
\end{figure}

Figure~\ref{FigA1}(a) shows the temperature dependence of the in- and out-of-plane magnetization at a magnetic field strength of $\mu_0H=0.1$T.
At this field strength, the calculated magnetization differs from the measurement in that the out-of-plane magnetization is larger than the in-plane one.
For $T\rightarrow0$ the in-plane magnetization is finite, whereas $M_{\perp}$ goes to zero.
This behavior is also found with an alternative model described in the next section, which better captures the magnetic properties at small magnetic fields.
Next, we show the temperature dependence of the magnetic susceptibility in Figure~\ref{FigA1}(b).
Again, the zero field curve deviates qualitatively from our measurements since it shows a small hump around $2-3$K.
At higher field strengths the qualitative behavior of the susceptibility is correctly captured, although the observed drop in susceptibility at $\mu_0H_{\parallel}=7.5$T above $T=3$K is very flat in our calculation.

\begin{figure}[]
	\includegraphics [width=0.9\linewidth, angle=0]{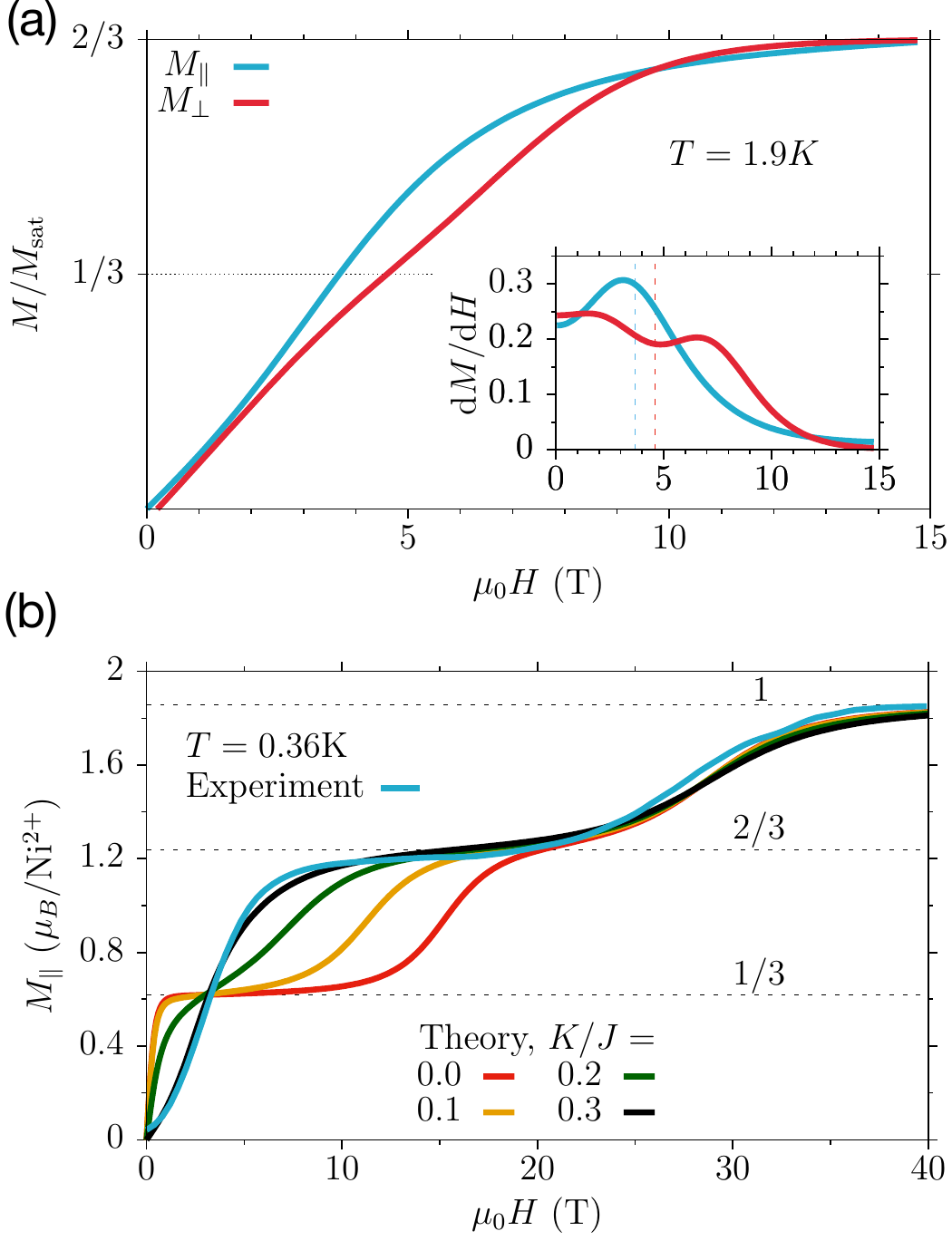}
	\caption{\footnotesize Field dependence of the magnetic properties of the model. (a) In- and out-of-plane magnetization at $T=1.9$K. The inset shows the derivative $\mathrm{d}M/\mathrm{d}H$ and the dashed lines correspond to the field strength for which $M/M_{\mathrm{sat}}=1/3$. (b) Effect of the strength of the biquadratic exchange term on the in-plane magnetization. All parameters of the model except of the biquadratic exchange $K/J$ are as described in the paper.}
	\label{FigA2}
\end{figure}

In Figure~\ref{FigA2}(a) the in- and out-of-plane magnetization is shown as a function of magnetic field strength $\mu_0H$ at $T=1.9$K.
As noted already in Fig.~\ref{FigA1}(a) the role of in- and out-of-plane magnetization is opposite to the measurements.
Both $M_{\parallel}$ and $M_{\perp}$ show a plateau at $2/3$ saturation magnetization and no $1/3$ plateau.
However, from the derivative $\mathrm{d}M/\mathrm{d}H$ shown in the inset one can read off a reduction in the slope of $M_{\perp}$ at the field value that corresponds to $1/3$ saturation magnetization indicating the remnant of a $1/3$ plateau.
The origin of this remnant feature becomes clearer when reducing the biquadratic exchange strength $K/J$ as shown in Figure~\ref{FigA2}(b).
For $K/J=0.2$ $M_{\parallel}$ shows a clear anomaly at $1/3 M_{\mathrm{sat}}$ and smaller values of $K/J$ give rise to a well-defined $1/3$ plateau.
From comparing the size of the $1/3$ and $2/3$ plateau the value of $K/J=0.3$ gives best agreement with experiment.

\section{Minimal Alternative Model}
\label{App:AltModel}
In the paper, we have presented a model that consists of three $S=1$ spins that are coupled via Heisenberg and biquadratic exchange and that experience a non-negligible Dzyaloshinskii-Moriya interaction.
This model accounts for the intriguing magnetization curve up to high magnetic field strengths and agrees well with the measured specific heat measurements of BHAP-Ni${}_3$.
However, details of the low-field behavior of magnetization and susceptibility are not fully captured.

Here, we investigate an alternative model, which takes into account only the Ni-Ni Heisenberg interactions and a single-ion anisotropy term.
Based on the Ni-O-Ni bonding angles being close to $90^{\circ}$, one would -in contrast to our DFT+U estimates- rather assume ferromagnetic exchange constants for two of the Ni-Ni exchange parameters.  
For one antiferromagnetic and two ferromagnetic exchange constants the triangular system is also magnetically frustrated and indeed shows a feeble $1/3$ and robust $2/3$ magnetization plateau. 

\begin{figure}[tb]
	\includegraphics [width=0.99\linewidth, angle=0]{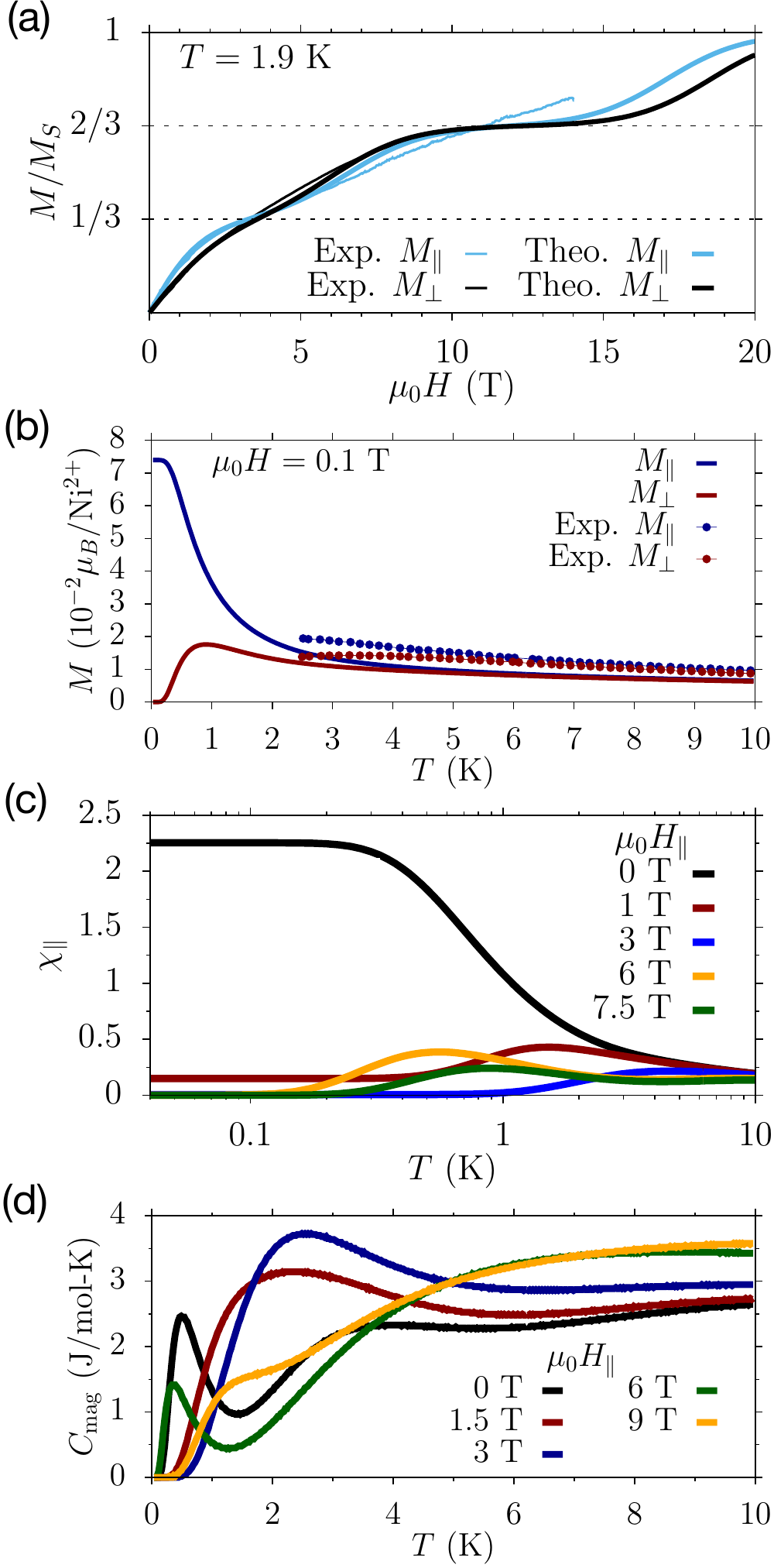}
	\caption{ Properties of the alternative model: Magnetization curve as a function of applied field strength (a) and temperature (b) for in-plane and perpendicular magnetic field direction. (c) In-plane susceptibility and (d) specific heat data as a function of temperature for different magnetic field strengths.}
	\label{NewMod}
\end{figure}

In the following, we present results for the minimalistic alternative model, described by:
$$\mathcal{H} = \sum_{\langle i,j\rangle}J_{ij}\vec{S}_i \cdot \vec{S}_j + D \sum_i \left(S_i^z \right)^2 + g\mu_B H\sum_i S_i^z.$$
Choosing the exchange parameters such that we reporduce the low-field magnetization data, we obtain $J_{12}=1.55 \mathrm{meV}, J_{23}=-1.29 \mathrm{meV}, J_{13}=-0.54 \mathrm{meV}.$
With a single-ion anisotropy of $D=0.12 \mathrm{meV}$ and the ESR g-factor of $g=2.23$ we obtain surprisingly good agreement with the in- and out-of-plane magnetization data at low magnetic fields $0\mathrm{ T}\leq H\lesssim 5\mathrm{ T}$, see Fig.~\ref{NewMod}(a).
At very low fields, this model also captures qualitatively the correct behavior of the magnetizations as a function of temperature [Fig.~\ref{NewMod}(b)].
The zero-field susceptibility is constant at very low temperatures [Fig.~\ref{NewMod}(c)], which is again consistent with experiment.

At small magnetic field strengths up to $3$T,  the in-plane susceptibility shows a temperature dependence, which slightly deviates from our measurements: Instead of a flat susceptibility at low temperatures up to a field strength of $3T$, the calculation shows a well-pronounced peak at $\sim 2$K already for field strengths of $1$T. 
Furthermore, the position of this peak varies notably with applied magnetic field strength, whereas the peak position remains rather constant in the measurement.

The most striking differences between this model and experiment occurs in the specific heat data. 
At zero magnetic field, the specific heat curve [Fig.~\ref{NewMod}(d)] shows a rather sharp peak at $\sim0.5$ T and a broader hump at $\sim3.5$ T.
The broad feature is seen in experiment, but the sharp peak disagrees with the measured specific heat data.
Also the increase in intensity for intermediate field strengths and the change of the position of the broad peak do not match the measurement.
Finally at $6$T there is a revival of the sharp peak feature at low temperature and even at $9$T the specific heat remains non-zero at fields larger than $\sim 0.5$T.

Overall, remarkably good agreement between experiment and this appealingly simple model is found for the magnetic properties of BHAP-Ni${}_3$ at low magnetic field strengths.
However, it is not capable of providing a satisfying description of both the intermediate and large-field magnetization curve and the thermal properties of BHAP-Ni${}_3$.
A model that could explain all salient features of the magnetization and thermal measurements of BHAP-Ni${}_3$ remains to be found.
This compound therefore presents an intriguing new challenge to theories for frustrated magnetism.

\bibliography{bhapni3}


\end{document}